\documentclass[12pt]{article}
\usepackage[latin1]{inputenc}
\usepackage{amsfonts}
\usepackage{amssymb}
\usepackage{graphics,psboxit,amsmath, epsfig}
\usepackage{psfrag}
\usepackage{verbatim}
\usepackage{fancyhdr}

\def\hybrid{\topmargin -20pt    \oddsidemargin 0pt
        \headheight 0pt \headsep 0pt
        \textwidth 6.35in       
        \textheight 9.25in       
        \marginparwidth .875in
        \parskip 5pt plus 1pt   \jot = 1.5ex}

\hybrid

\def\baselinestretch{1.2}

\catcode`\@=11
\def\marginnote#1{}
\newcount\hour
\newcount\minute
\newtoks\amorpm
\hour=\time\divide\hour by60
\minute=\time{\multiply\hour by60 \global\advance\minute by-\hour}
\edef\standardtime{{\ifnum\hour<12 \global\amorpm={am}%
        \else\global\amorpm={pm}\advance\hour by-12 \fi
        \ifnum\hour=0 \hour=12 \fi
        \number\hour:\ifnum\minute<10 0\fi\number\minute\the\amorpm}}
\edef\militarytime{\number\hour:\ifnum\minute<10 0\fi\number\minute}

\def\draftlabel#1{{\@bsphack\if@filesw {\let\thepage\relax
   \xdef\@gtempa{\write\@auxout{\string
      \newlabel{#1}{{\@currentlabel}{\thepage}}}}}\@gtempa
   \if@nobreak \ifvmode\nobreak\fi\fi\fi\@esphack}
        \gdef\@eqnlabel{#1}}
\def\@eqnlabel{}
\def\@vacuum{}
\def\draftmarginnote#1{\marginpar{\raggedright\scriptsize\tt#1}}

\def\draft{\oddsidemargin -.5truein
        \def\@oddfoot{\sl preliminary draft \hfil
        \rm\thepage\hfil\sl\today\quad\militarytime}
        \let\@evenfoot\@oddfoot \overfullrule 3pt
        \let\label=\draftlabel
        \let\marginnote=\draftmarginnote
   \def\@eqnnum{(\theequation)\rlap{\kern\marginparsep\tt\@eqnlabel}%
\global\let\@eqnlabel\@vacuum}  }

\def\preprint{\twocolumn\sloppy\flushbottom\parindent 2em
        \leftmargini 2em\leftmarginv .5em\leftmarginvi .5em
        \oddsidemargin -.5in    \evensidemargin -.5in
        \columnsep .4in \footheight 0pt
        \textwidth 10.in        \topmargin  -.4in
        \headheight 12pt \topskip .4in
        \textheight 6.9in \footskip 0pt
        \def\@oddhead{\thepage\hfil\addtocounter{page}{1}\thepage}
        \let\@evenhead\@oddhead \def\@oddfoot{} \def\@evenfoot{} }

\def\numberbysection{\@addtoreset{equation}{section}
        \def\theequation{\thesection.\arabic{equation}}}

\def\underline#1{\relax\ifmmode\@@underline#1\else
        $\@@underline{\hbox{#1}}$\relax\fi}

\def\titlepage{\@restonecolfalse\if@twocolumn\@restonecoltrue\onecolumn
     \else \newpage \fi \thispagestyle{empty}\c@page\z@
        \def\thefootnote{\fnsymbol{footnote}} }

\def\endtitlepage{\if@restonecol\twocolumn \else \newpage \fi
        \def\thefootnote{\arabic{footnote}}
        \setcounter{footnote}{0}}  

\catcode`@=12
\relax

\def\figcap{\section*{Figure Captions\markboth
        {FIGURECAPTIONS}{FIGURECAPTIONS}}\list
        {Figure \arabic{enumi}:\hfill}{\settowidth\labelwidth{Figure
999:}
        \leftmargin\labelwidth
        \advance\leftmargin\labelsep\usecounter{enumi}}}
 \relax
\def\tablecap{\section*{Table Captions\markboth
        {TABLECAPTIONS}{TABLECAPTIONS}}\list
        {Table \arabic{enumi}:\hfill}{\settowidth\labelwidth{Table
999:}
        \leftmargin\labelwidth
        \advance\leftmargin\labelsep\usecounter{enumi}}}
 \relax
\def\reflist{\section*{References\markboth
        {REFLIST}{REFLIST}}\list
        {[\arabic{enumi}]\hfill}{\settowidth\labelwidth{[999]}
        \leftmargin\labelwidth
        \advance\leftmargin\labelsep\usecounter{enumi}}}
 \relax
\makeatletter
\newcounter{pubctr}
\def\publist{\@ifnextchar[{\@publist}{\@@publist}}
\def\@publist[#1]{\list
        {[\arabic{pubctr}]\hfill}{\settowidth\labelwidth{[999]}
        \leftmargin\labelwidth
        \advance\leftmargin\labelsep
        \@nmbrlisttrue\def\@listctr{pubctr}
        \setcounter{pubctr}{#1}\addtocounter{pubctr}{-1}}}
\def\@@publist{\list
        {[\arabic{pubctr}]\hfill}{\settowidth\labelwidth{[999]}
        \leftmargin\labelwidth
        \advance\leftmargin\labelsep
        \@nmbrlisttrue\def\@listctr{pubctr}}}
 \relax
\makeatother
\newskip\humongous \humongous=0pt plus 1000pt minus 1000pt

\newif\ifdtup

\relax

\def\be{\begin{equation}}
\def\ee{\end{equation}}
\def\ba{\begin{eqnarray}}
\def\ea{\end{eqnarray}}

\def\IR{\relax{\rm I\kern-.18em R}}
\def\II{\relax{\rm 1\kern-.35em1}}

\renewcommand{\theequation}{\thesection.\arabic{equation}}
\csname @addtoreset\endcsname{equation}{section}

\def\N4{${\cal N}=4$}

\def\bo{\boldsymbol }

\newcommand{\Feyn}[1]{#1\kern-0.45em/}

\begin{document}

\title{\Large \bf Dual conformal invariance in the Regge limit}
\author{\large C{\'e}sar~G{\'o}mez, Johan~Gunnesson, Agust{\'i}n~Sabio~Vera \\
{\it Instituto de F{\' i}sica Te{\' o}rica UAM/CSIC,}\\ 
{\it Universidad Aut{\' o}noma de Madrid, E-28049 Madrid, Spain}}

\maketitle

\vspace{-9cm}
\begin{flushright}
{\small IFT--UAM/CSIC--09--38}
\end{flushright}

\vspace{7cm}
\begin{abstract}
\noindent

A dual conformal symmetry, analogous to the dual conformal symmetry observed for the scattering amplitudes of \N4 Super Yang-Mills theory, is identified in the Regge limit of QCD. Combined with the original two-dimensional conformal symmetry of the theory, this dual symmetry can potentially explain the integrability of the BFKL Hamiltonian. We also give evidence that the symmetry survives when a subset of unitarity corrections are taken into account by studying briefly the non-planar $2$ to $m$ reggeon transition vertices.
   
\end{abstract}

\section{Introduction}

In the last few years there has been a great deal of progress in the study of gluon scattering amplitudes in 
the maximally supersymmetric gauge theory in four dimensions, \N4 Super Yang-Mills (SYM). One of the most surprising 
developments has been the discovery of a hidden symmetry in the planar ($N_c \rightarrow \infty$) limit, coined 
as ``dual super-conformal symmetry''~\cite{dualconformal, korchemskyconf,confward}, different from the original 
super-conformal symmetry of the Lagrangian. This symmetry was uncovered by introducing a new set of variables 
$x_i$, related to the external (all taken as incoming) gluon momenta $p_i$, $i=1\ldots n$, through
\be
x_i - x_{i+1}=p_i \ , \label{eq:xi}
\ee
and acts on the $x_i$ just as a four-dimensional conformal symmetry acts on spatial coordinates. The presence 
of this dual symmetry can be understood through the AdS/CFT correspondence~\cite{AdSCFT} since it was 
shown~\cite{fermionicT} that the problem of calculating a given scattering amplitude can be mapped, 
through a fermionic T-duality, to that of calculating a light-like Wilson loop with corners at coordinates 
given by the $x_i$. This fermionic T-duality maps the string $\sigma$-model to itself, and the dual conformal 
symmetry becomes the ordinary symmetry of the space in which the Wilson loop lives. 

Like the ordinary conformal symmetry, the dual symmetry is broken by infrared divergences, arising as cusp 
divergences in the language of Wilson loops. However, the cusp divergences are known to exponentiate, which 
allows the use of the broken symmetry to impose powerful constraints on the amplitudes in the form of anomalous 
Ward identities. These identities fix the 4 and 5 point amplitudes completely while the undetermined parts of 
higher-point amplitudes can only depend on dual-conformal invariants. Also, taken together, the original and 
dual conformal symmetries generate an infinite-dimensional Yangian symmetry~\cite{plefka}, ordinarily characteristic 
of exactly solvable models.

Interesting properties of \N4 amplitudes appear also in their high energy (Regge) limit. In a nutshell, Regge theory 
establishes the structure of scattering amplitudes when the momentum transfer is small compared to the total 
center-of-mass energy. It turns out that \N4 amplitudes exhibit Regge-like behaviour at all orders in the 't 
Hooft coupling even outside of the Regge limit. In fact, the 4 and 5 point amplitudes are Regge 
exact~\cite{korchemskyconf, AgustinBartelsLipatov}, 
meaning that they can always be written in a factorized form characteristic of high energies, irrespective of 
the values of the kinematical invariants\footnote{Also, a proposal for the undetermined part of the 6 point amplitude, having the correct Regge behaviour and given in terms of conformal cross-ratios, is given in the latest version of \cite{AgustinBartelsLipatov2}. }. Furthermore, in the Leading Logarithmic Approximation (LLA) of gluon 
amplitudes the Regge limit is independent of the gauge theory, so \N4 can give insight into the high energy 
behaviour of QCD.

In the Regge limit amplitudes are dominated by the $t$-channel exchange of reggeized gluons (a reggeized gluon is 
a collective state of ordinary gluons projected on a colour octet). The bound state of two reggeized gluons when 
projected on a colour singlet in the $t$-channel is known as the hard (or perturbative) pomeron. The interaction between 
reggeized gluons is governed by the Schr{\"o}dinger-like BFKL integral equation~\cite{BFKL} where the invariant mass of 
$s$-channel gluons can be interpreted as the time variable and its kernel as an effective Hamiltonian living on the 
two-dimensional transverse space. This Hamiltonian is free from infrared singularities and carries a yet to be understood 
integrability~\cite{BFKLintegrable} \footnote{Integrability also appears when the gluon composite states are projected onto the
adjoint representation~\cite{Lipatov:2009nt}.}.

The question that arises is if the integrable structures present in \N4 SYM can shed some light on the integrability 
found in the Regge limit. In this region the dynamics of the theory is reduced to the transverse plane, where a two-dimensional 
conformal symmetry was found in the effective Hamiltonian~\cite{Lipatov1986}. Given the emergence of the Yangian in the 
four-dimensional case, and that one can heuristically interpret this $SL(2,C)$ as a reduction of the four-dimensional ordinary conformal 
symmetry, it would then seem natural to look for a dual $SL(2,C)$ symmetry in the high-energy limit\footnote{Hints in this direction have already appeared in the literature. A similar dual symmetry was exploited in \cite{Lipatov:2009nt} in order to map supersymmetric multiparticle amplitudes in multiregge kinematics to an integrable open spin chain. In fact, the octet kernel, after subtraction of infrared divergences, can be written in a form manifestly invariant under this symmetry. Also, the BFKL Hamiltonian has holomorphic separability into two pieces which can be written such that they are invariant under the duality transformation $p_i \to \rho_i - \rho_{i+1}
\to p_{i+1}$, similar to the change of variables \eqref{eq:xi}, with the $\rho$ being the gluon transverse coordinates in
complex notation~\cite{Lipatov:1998as}.}. It is this question that we address in this Letter, showing that BFKL indeed exhibits covariance under such a dual symmetry.

We will also study a set of corrections to BFKL, in the form of $2\rightarrow m$ reggeized gluon transition 
vertices, which also turn out to be dual $SL(2,C)$-covariant. This result is important for high-energy QCD, since the inclusion of such vertices is necessary, at sufficiently high energies, to fulfill unitarity in all channels.

\section{The dual $SL(2,C)$ symmetry}

\begin{figure}[ht]
\psfrag{+}{$+$} \psfrag{=}{$=$} \psfrag{ka}{$k_A$} \psfrag{kb}{$k_B$}
\psfrag{kp}{$k'$} \psfrag{F}{$F$} \psfrag{kamq}{$k_A - q$}
\psfrag{kbmq}{$k_B - q$} \psfrag{x1}{$x_1$} \psfrag{x2}{$x_2$}
\psfrag{x3}{$x_3$} \psfrag{x4}{$x_4$}
\begin{center}
\includegraphics{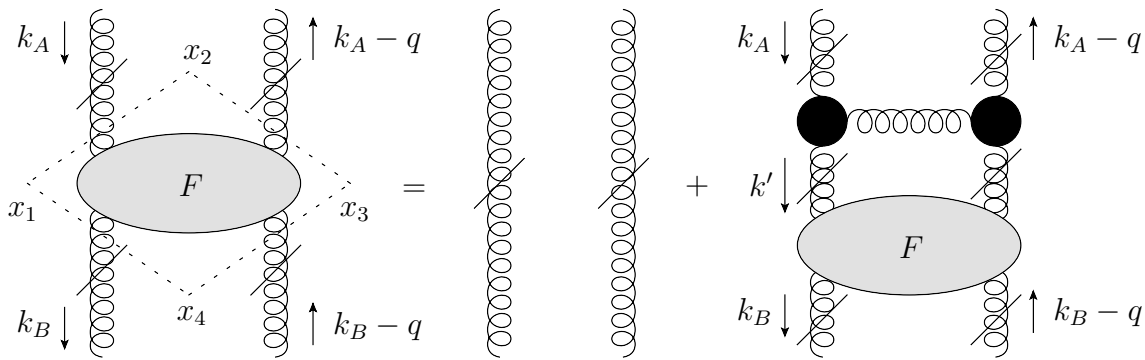}
\caption{\small The BFKL integral equation for the four-point reggeized gluon
Green function.} \label{fig:BFKL}
\end{center}
\end{figure}

The scattering amplitude for the 2 to 2 reggeized gluons process in the Regge limit has an iterative structure 
dominated by the exchange in the $t$-channel of a colour singlet. This implies that in the LLA the corresponding 
4 point gluon Green function can be written as the solution to an integral 
equation, the BFKL equation \footnote{For an introductory treatment of the BFKL equation, see \cite{rossforshaw}}, shown in Fig.~\ref{fig:BFKL}. Written in terms of $\omega$, the Mellin conjugate variable of the 
center-of-mass energy (which can be translated into 
the rapidity, $Y$, of the emitted particles in the $s$-channel)), and the incoming two dimensional momenta it reads   
\begin{equation}
\omega F(\omega , \, \bo{k}_A,\, \bo{k}_B,\, \bo{q}) = \delta^{(2)}(\bo{k}_A - \bo{k}_B) + \int d^2\bo{k}' K(\bo{k}_A,\,
 \bo{k}_A-\bo{q};\, \bo{k}',\, \bo{k}'-\bo{q}) F(\omega , \, \bo{k}',\, \bo{k}_B ,\, \bo{q}) \, \label{eq:BFKLconK} 
\end{equation}
where the kernel $K(\bo{k}_A,\, \bo{k}_A-\bo{q};\, \bo{k}',\, \bo{k}'-\bo{q})$ is given by
\begin{align}
\frac{K_R (\bo{k}_A,\, \bo{k}_A-\bo{q}; -\bo{k}'+\bo{q}, \, -\bo{k}')}{8\pi^3\bo{k}_A^2(\bo{k}' - \bo{q})^2} 
+ \left[  \omega (\bo{k}_A^2) + \omega ((\bo{k}_A-\bo{q})^2) \right]\delta ^{(2)}(\bo{k}_A-\bo{k}') \ . \label{eq:Kernel}
\end{align}
The ``real emission'' part\footnote{Due to the optical theorem when the forward ${\bf q}=0$ limit is taken this piece in 
the kernel corresponds to the contribution to multiparticle production from on-shell gluons in the $s$-channel.} 
has the following structure
\be
K_R (\bo{p}_1,\, \bo{p}_2;\, \bo{p}_3,\, \bo{p}_4) = -N_cg^2\left[ \left( \bo{p_3}+\bo{p_4} \right)^2 - \frac{\bo{p}_2^2\bo{p}_4^2}{(\bo{p}_2+\bo{p}_3)^2}- \frac{\bo{p}_1^2\bo{p}_3^2}{(\bo{p}_1+\bo{p}_4)^2}  \right]. \label{eq:KernelR}
\ee
This notation, with $p_1,\, \ldots ,\, p_4$ being the cyclically ordered reggeized gluon momenta taken as incoming, 
will be convenient for the generalization of this vertex to the $2 \rightarrow m$ reggeized gluon transition case 
as we will see below. 

The gluon Regge trajectory reads
\be
\omega (\bo{q}^2) = -\frac{g^2N_c}{16 \pi ^3 }\int d^2\bo{k}' \frac{\bo{q}^2}{\bo{k}'^2(\bo{k}'-\bo{q})^2} \ . 
\label{eq:trajectoryreggegluon}
\ee
The trajectory is IR divergent, requiring it to be regularized in general.

We will now show that the BFKL equation in Eq.~\eqref{eq:BFKLconK} exhibits formally a dual $SL(2,C)$ symmetry, which, 
in contrast with the original $SL(2,C)$ symmetry of BFKL, uncovered by Fourier transforming into a coordinate 
representation, is realized in the transverse momentum space. This new symmetry is closely analogous to the dual 
conformal symmetry observed in $\mathcal{N}=4$ SYM for gluon scattering amplitudes, and we will see that it turns out 
to be broken by infrared effects just as in the four dimensional gauge theory.

Let us now rewrite Eq.~\eqref{eq:BFKLconK} in terms of dual variables. Taken as incoming, the external momenta are 
$\bo{k}_A$, $-\bo{k}_A+ \bo{q}$, $\bo{k}_B - \bo{q}$ and $-\bo{k}_B$ so, introducing the notation 
$x_{i,j}\equiv x_i - x_j$, we define the new set of variables as
\begin{equation}
\bo{p}_1 = x_{1,2} = \bo{k}_A, \, \bo{p}_2 = x_{2,3} = \bo{q} -\bo{k}_A, \,
\bo{p}_3 = x_{3,4} = \bo{k}_B - \bo{q}, \, \bo{p}_4=x_{4,1} = -\bo{k}_B. 
\end{equation}
Equivalently, we could have written $\bo{k}_A = x_{1,2}, \, \bo{k}_B = x_{1,4}, \, 
\bo{q} = x_{1,3}$ with $x_1$ then being a simple shift of the origin for the external momenta.

In these new variables the gluon Regge trajectory is
\be
\omega (\bo{k}_A^2) = \omega (x^2_{1,2})=-\frac{g^2N_c}{16 \pi ^3 }\int d^2 x_I \frac{x^2_{1,2}}{x^2_{I,1}x^2_{I,2}} \ , \label{eq:trajectoryreggegluonk1x}
\ee
where we have introduced $x_I$ through $\bo{k}' = x_{I,2}$. Ignoring for the moment that this expression is divergent and thus ill-defined, we see that is has a formal two-dimensional conformal symmetry. It is formally invariant under translations, rotations and scalings of the $x_i$, and also under the conformal inversions $x_i \rightarrow \frac{x_i}{x_i^2}$, since they would imply
\be
d^2 x_I \rightarrow \frac{d^2 x_I}{x_I^4} \ , \;\; x_{i,j}^2 \rightarrow \frac{x_{i,j}^2}{x_i^2x_j^2} \ . \label{eq:transfxij} 
\ee
In the same way $\omega ((\bo{k}_B-\bo{q})^2) = \omega (x^2_{2,3})$ is also formally conformally invariant. 
Now, given that the trajectory is infrared divergent one would expect this symmetry to be broken by the introduction 
of a regulator, an issue which is discussed in the next section.

Rewriting the full kernel~\eqref{eq:Kernel} in terms of the $x_i$, with $\bo{k}'=x_{1,I}$, we get
\be
K(x_{1,2},\, x_{3,2};\, x_{1,I},\, x_{3,I}) = 
\frac{K_R (x_{1,2},\, x_{2,3}; x_{3,I}, \, x_{I,1})}{8\pi^3x_{1,2}^2x_{I,3}^2} 
+ \left[  \omega (x_{1,2}^2) + \omega (x_{2,3}^2) \right]\delta ^{(2)}(x_{2,I}) \ , \label{eq:Kernelx}
\ee
where
\be
K_R (x_{1,2},\, x_{2,3}; x_{3,I}, \, x_{I,1}) = -N_cg^2\left[  x_{1,3}^2 - \frac{x_{2,3}^2 x_{I,1}^2}{x_{2,I}^2}- \frac{x_{1,2}^2x_{I,3}^2}{x_{2,I}^2}  \right]  \ . \label{eq:KernelRx}
\ee
Using that $\delta ^{(2)}(x_{2,I}) \rightarrow x_2^2x_I^2 \delta ^{(2)}(x_{2,I})$ under conformal inversions one then finds immediately that the kernel transforms covariantly \footnote{This is again similar to the (conjectured) dual conformal symmetry of scattering amplitudes in \N4 SYM, under which the amplitudes transform covariantly, as opposed to the ordinary conformal symmetry which leave them invariant.}
\be
K(x_{1,2},\, x_{3,2};\, x_{1,I},\, x_{3,I}) \rightarrow  x^2_2x^2_I K(x_{1,2},\, x_{3,2};\, x_{1,I},\, x_{3,I}) \ .
\ee
Together with translations, rotations and dilatations this forms a dual $SL(2,C)$ symmetry, different from the one 
previously known. More precisely, dilatations and rotations coincide with the original $SL(2,C)$-symmetry, while 
translations and inversions will be different. 

Applied to the BFKL equation~\eqref{eq:BFKLconK} and using that the integration measure transforms according 
to~\eqref{eq:transfxij} one finds that a factor of $\frac{x_2^2}{x_I^2}$ is produced inside the integral. 
Consequently, if the Green function $F(\omega , \, x_{1,2},\, x_{1,4},\, x_{1,3})$ were to produce a factor of 
$x_2^2$ upon inversion, then its convolution with the kernel $K \otimes F$, would transform in the same way as $F$ 
itself. Now, at lowest order, $F$ is simply given by the delta function, which indeed transforms in this way, 
$\delta^{(2)} (\bo{k}_A - \bo{k}_B)=\delta ^{(2)} (x_{2,4}) \rightarrow x_2^2 x_4^2 \delta ^{(2)} (x_{2,4})$. 
Since $F$ can be constructed through iterated convolution with the kernel, it follows that the Green function 
should have the same conformal properties as the delta function. 

We can obtain a formal expression for the Green function having the correct conformal properties by iteration. 
Introducing the short-hand notation
\be
\omega_0 \left(\bf{k}_A,\, \bf{q} \right) \equiv \omega (\bo{k}_A^2) + \omega ((\bo{k}_A-\bo{q})^2),  
\xi \left(\bf{k},\, \bf{k}_A,\, \bf{q}\right) \equiv 
\frac{K_R (\bo{k}_A,\, \bo{k}_A-\bo{q}; -\bo{k}+\bo{q}, \, -\bo{k})}{8\pi^3 \bo{k}_A^2(\bo{k} - \bo{q})^2} \ 
\ee
one finds (with $\bf{k}_0 \equiv \bf{k}_A$):
\begin{eqnarray}
F \left(\omega,\bf{k}_A,\bf{k}_B,\bf{q}\right) =
{\delta^{(2)} \left(\bf{k}_A-\bf{k}_B\right) + \sum_{n=1}^\infty \prod_{i=1}^n 
\int d^2 \bf{k}_i \, {\xi \left(\bf{k}_i,\bf{k}_{i-1},\bf{q}\right) \over 
\omega - \omega_0 \left(\bf{k}_i,\bf{q}\right)} \delta^{(2)}
\left(\bf{k}_n-\bf{k}_B\right) \over \omega - \omega_0 \left(\bf{k}_A,\bf{q}\right)}.  
\end{eqnarray}

Rather than $\omega$ it is more natural to use the rapidity difference, $Y$, between the external 
particles as the evolution variable. To this end we perform the inverse Mellin transform
\begin{eqnarray}
{\cal F} \left(\bf{k}_A,\bf{k}_B,\bf{q},Y\right) &=&
\int_{a-i \infty}^{a+i \infty} {d \omega \over 2 \pi i} e^{\omega Y}
F \left(\omega,\bf{k}_A,\bf{k}_B,\bf{q}\right).
\end{eqnarray}
The formula $\int_{a-i \infty}^{a+i \infty} {d \omega \over 2 \pi i} e^{\omega Y}
\prod_{i=0}^n \frac{1}{\omega-\omega_i} = e^{\omega_0 Y} \prod_{i=1}^n
\int_0^{y_{i-1}} d y_i e^{\omega_{i,i-1} y_i}$ for $n > 0$, with $\omega_{i,j} \equiv 
\omega_i - \omega_j, y_0 \equiv Y$, is useful to obtain the final expression, written in dual $x$-variables:
\begin{equation}
{\cal F} \left(x_{12}, x_{14},x_{13},Y\right) = e^{\omega _{2,1} Y} \Bigg\{\delta^{(2)} \left(x_{24}\right)
+ \sum_{n=1}^\infty \prod_{i=1}^n \int_0^{y_{i-1}} \hspace{-0.3cm} d y_i
\int d^2 x_i \, \xi_{i,i-1} e^{\omega_{i,i-1} y_i} \delta^{(2)}
\left(x_{4,n} \right) \Bigg\}, 
\label{eq:Frapidityx}
\end{equation}
where
\begin{eqnarray}
\omega_{i,i-1} &=& \omega _0 \left(x_{1,i},x_{13}\right)-\omega_0 \left(x_{1,i-1},x_{13}\right) \ ,\\
\xi_{i,i-1} &=& {{\bar \alpha}_s \over 2 \pi} \Bigg\{{ x^2_{i,3} x^2_{i-1,1} +
x^2_{i-1,3} x^2_{i,1} - x^2_{i-1,i} x^2_{13}\over x^2_{i,3} x^2_{i-1,i} x^2_{i-1,1}}\Bigg\} \ .
\end{eqnarray}
This representation preserves the transformation properties of the original equation. In the forward case, where the 
momentum transfer is zero, the same structure remains with  
\be
\omega_{i,i-1} = 2\left(\omega \left(x_{1,i}\right)-\omega \left(x_{1,i-1}\right)\right) \ ,
\xi_{i,i-1} = {{\bar \alpha}_s \over \pi} { 1 \over x^2_{i-1,i}} \ .
\ee
In this case the solution also has a formal dual $SL(2,C)$ covariance. This should be contrasted with 
the original $SL(2,C)$-invariance of the BFKL kernel, which does not appear in the forward case.

Before ending this section, it is noteworthy to mention that this formal dual $SL(2,C)$ covariance is 
present in the same form for all color projections in the $t$-channel since they only differ by a different factor in front of $K_R$: with $N_c=3$, $c_1 = 1,\, c_{8_a} = c_{8_s} = 1/2,\, c_{10} = c_{\overline{10}} = 0$, etc.

\section{The effect of IR divergences}
 
In \N4 SYM infrared divergences break the dual conformal symmetry. For BFKL, such divergences cancel, opening the 
possibility that the dual $SL(2,C)$-symmetry remains exact. However, this turns out not to be the case. Perhaps the 
simplest way to see this is by studying the forward case. If $F$ has the transformation properties of the delta 
function it can be written as
\be
F = F_1 \delta ^{(2)}(\bo{k}_A - \bo{k}_B) + \frac{1}{(\bo{k}_A - \bo{k}_B)^2}F_2 \ ,
\ee
where $F_1$ and $F_2$ are dual conformally invariant, since $(\bo{k}_A - \bo{k}_B)^{-2}$ is the only other function 
that transforms correctly. When $\bo{q}=0$, $x_1=x_3$ and no non-trivial conformal invariant can be formed from the 
three remaining $x_i$. $F_2$ can thus only be a function of $\omega$ (or equivalently the rapidity $Y$), and the 
coupling. But when forming physical quantities one integrates over $\bo{k}_A$ and $\bo{k}_B$ and the divergences 
at $\bo{k}_A=\bo{k}_B$ must cancel between $F_1$ and $F_2$. The factor $(\bo{k}_A - \bo{k}_B)^{-2}$ is singular 
enough to cancel one factor of the trajectory, but $F_1$ is obtained by repeated application of the trajectory 
part of the kernel so, starting from the second iteration, products of two or more trajectories will appear and 
the divergences will fail to cancel.

One can also observe the breakdown of the dual $SL(2,C)$ symmetry directly by regularizing the integrals and 
cancelling the divergences explicitly when performing the iteration. One then finds that the first iteration 
respects the symmetry, while the second iteration produces a contribution to $F_2$ (when $\bo{q}=0$) proportional 
to an anomalous factor of the form $\ln \left( \frac{(\bo{k}_A-\bo{k}_B)^4}{\bo{k}_A^2 \bo{k}_B^2}\right)$, which 
breaks the symmetry under inversions. The origin of this factor is the regularization of infrared divergences. 
For example, using dimensional regularization with $D=4-2\epsilon$
\be
\omega (x^2_{1,2})=-\frac{g^2N_c}{16 \pi ^3 }(4 \pi \mu)^{2\epsilon}\int d^{2-2\epsilon} x_I \frac{x^2_{1,2}}{x^2_{I,1}x^2_{I,2}}\approx -\frac{g^2N_c}{8 \pi ^2 }(4\pi e^{-\gamma})^\epsilon  \left( \ln \frac{x^2_{1,2}}{\mu ^2}- \frac{1}{\epsilon} \right) \ .
\ee
The divergences will cancel between the trajectories and the real emission part of the kernel, but factors such 
as $\ln x^2_{1,2}$ will add up giving a non-vanishing anomalous term. So, even though BFKL is infra-red finite, 
a remnant of the divergences remains in the form of the breaking of the dual $SL(2,C)$ symmetry in the form given here.

Further insight can be gained by studying a standard representation of the Green function in the forward case, 
obtained by diagonalizing the BFKL kernel. It is
\be
{\cal F} \left(x_{1,2},x_{1,4},Y\right) = 
\sum_{n=-\infty}^\infty \int \frac{d \gamma}{2 \pi i}
\left({x^2_{1,2} \over x^2_{1,4}}\right)^{\gamma-\frac{1}{2}}
\frac{e^{{\bar{\alpha}_s}\chi_n\left(\gamma\right) Y + i n \theta_{2,4}}}{\pi\sqrt{x_{1,2}^2 x_{1,4}^2}}\ ,
\ee
with $\chi_n  \left(\gamma\right) =  2 \Psi\left(1\right)- \Psi\left(\gamma+\frac{|n|}{2}\right)
- \Psi\left(1-\gamma+\frac{|n|}{2}\right)$ and 
$\cos{\theta_{2,4}} = { x_{1,2} \cdot x_{1,4} \over \sqrt{x^2_{1,2} x^2_{1,4}}}$.
In this representation any dependence on an IR cutoff has canceled explicitly, and one can check that the covariance under conformal inversions is lost.

An important issue is whether the dual $SL(2,C)$ symmetry is broken beyond repair or whether it can be deformed to take into consideration the anomalous terms. In a best case scenario the symmetry would obey to all orders a simple relation such as the anomalous Ward identity satisfied by the dual conformal symmetry of $\mathcal{N}=4$ scattering amplitudes \cite{confward}. This issue is studied in \cite{johan}, with the result that the dual $SL(2,C)$ does not obey such a simple all-order relation, but can still be deformed so that it becomes exact, at least up to the order studied. The representation then becomes coupling-dependent, but encouragingly, it seems to do so in such a way that the algebra generated by the original and dual $SL(2,C)$ symmetries remains coupling-independent.

\section{$2\rightarrow m$ reggeized gluon vertex}

\begin{figure}[ht]
\psfrag{x1}{$x_1$} \psfrag{x2}{$x_2$} \psfrag{x3}{$x_3$} \psfrag{x4}{$x_4$} \psfrag{x5}{$x_5$} \psfrag{p1}{$\bo{p}_1$} \psfrag{p2}{$\bo{p}_2$} \psfrag{p3}{$\bo{p}_3$} \psfrag{p4}{$\bo{p}_4$} \psfrag{p5}{$\bo{p}_5$} \psfrag{pmp2}{$\bo{p}_{m+2}$} \psfrag{a1}{$a_1$} \psfrag{a2}{$a_2$}  \psfrag{amm1}{$a_{m-1}$} \psfrag{am}{$a_m$} \psfrag{b1}{$b_1$} \psfrag{b2}{$b_2$}
\begin{center}
\includegraphics{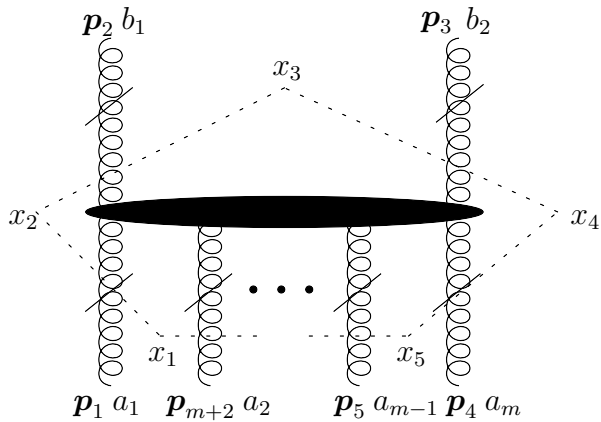}
\caption{\small The $2\rightarrow m$ reggeized gluon vertex. All momenta are taken as ingoing.} \label{fig:vertex}
\end{center}
\end{figure}

The BFKL amplitude will violate bounds imposed by unitarity at sufficiently high energies. 
In order to restore unitarity, one of the new elements that must be introduced is a vertex in which the number 
of reggeized gluons in the $t$-channel is not conserved. As shown in Fig.~\ref{fig:vertex} we choose to write this 
$2\rightarrow m$ vertex (see, for example, Eq.~(3.57) of \cite{BartelsEwerz}) using a convenient assignment 
of the momentum indices
\begin{align}
&K_{2\rightarrow m}^{\{ b \} \rightarrow \{a \}}(\bo{p}_2,\, \bo{p}_3;\, \bo{p}_4,\, \ldots ,\,  \bo{p}_{m+2},\, \bo{p}_1 ) = f_{a_1b_1c_1}f_{c_1a_2c_2}\cdots f_{c_{m-1}a_mb_2} g^m \nonumber \\
& \times 
\left[ (\bo{p}_4 + \cdots + \bo{p}_1)^2 - \frac{\bo{p}_3^2(\bo{p}_5 + \cdots + \bo{p}_1)^2}{(\bo{p}_3 + \bo{p}_4)^2}
 -  \frac{\bo{p}_2^2(\bo{p}_4 + \cdots + \bo{p}_{m+2})^2}{(\bo{p}_1 + \bo{p}_2)^2}  + \frac{\bo{p}_1^2\bo{p}_3^2(\bo{p}_5 + \cdots + \bo{p}_{m+2})^2}{(\bo{p}_1 + \bo{p}_2)^2(\bo{p}_3 + \bo{p}_4)^2} \right] \label{eq:K2m} \ ,
\end{align}
where the $a_1,\, b_1$ etc. are the color indices of the reggeized gluons and $f_{ijk}$ the structure constants of 
$SU(N_c)$. 

Written in terms of $x$ variables this becomes
\begin{align}
&K_{2\rightarrow m}^{\{ b \} \rightarrow \{a \}}(x_{23},\, x_{34};\, x_{45},\, \ldots ,\,  x_{m+2,1},\, x_{12}) = \nonumber \\ &f_{a_1b_1c_1}f_{c_1a_2c_2}\cdots f_{c_{m-1}a_mb_2} g^m \left[ x_{24}^2 - \frac{x_{34}^2x_{25}^2}{x_{35}^2} -\frac{x_{23}^2x_{14}^2}{x_{13}^2} +\frac{x_{23}^2x_{34}^2x_{15}^2}{x_{13}^2 x_{35}^2} \right] \ ,
\end{align}
and is manifestly conformally covariant. The assignment of the momenta in \eqref{eq:K2m} was chosen so that 
the vertex takes a form independent of $m$ when written in terms of the $x_i$. Note that the last term vanishes 
when $m=2$ since then $x_1 = x_5$, and one recovers the corresponding term in the BFKL kernel. 

In \cite{reggevertex2a4} it was shown that the $2\rightarrow 4$ reggeized gluon vertex exhibited the same coordinate representation $SL(2,C)$-invariance as the BFKL equation. This was taken to indicate that a unitary, two-dimensional CFT describing scattering amplitudes in the Regge limit should exhibit this $SL(2,C)$-invariance. Our results would seem to indicate that such a theory should also be covariant under the dual $SL(2,C)$.

\section{Conclusions}

We have shown that not only does the LLA BFKL kernel, and its extension in the form of the $2 \rightarrow m$ reggeized gluon vertex, exhibit the ordinary $SL(2,C)$-symmetry, found by Lipatov but also a dual $SL(2,C)$, analogous to the dual conformal symmetry of $\mathcal{N}=4$. It is tempting to interpret these symmetries as reductions to the transverse plane of the conformal and dual conformal symmetries of the supersymmetric theory, although it is not clear exactly how such a reduction should be carried out. Purely transverse versions of the conformal algebras are not symmetries of the 4-dimensional gauge theory amplitudes, but seem to emerge in the Regge limit. 

Also, the dual invariance of the reggeized gluon vertex suggests that a unitary two-dimensional CFT describing high-energy gauge theory should have both $SL(2,C)$ groups. In future work, having identified the dual $SL(2,C)$ one can try to understand the origin of the integrability of the Regge limit in terms of the integrability of $N=4$ SYM.


\vspace{5mm}
\centerline{\bf Acknowledgments}

We would like to thank Lev Lipatov for useful discussions. The work of C. G. has been partially
supported by the Spanish DGI contract FPA2003-02877 and the CAM grant HEPHACOS
P-ESP-00346. The work of J. G. is supported by a Spanish FPU grant.



\end{document}